\documentclass[11pt,a4paper]{article}
\usepackage{a4wide, axodraw}
\usepackage{cite}
\usepackage{amsmath,amsthm,amssymb,amsfonts}
\usepackage{array}
\usepackage{verbatim}
\makeatletter
\renewcommand\section{\@startsection {section}{1}{\z@}%
                                   {-3.5ex \@plus -1ex \@minus -.2ex}%nn
                                   {2.3ex \@plus.2ex}%
                                   {\normalfont\large\bfseries}}
\renewcommand\subsection{\@startsection{subsection}{2}{\z@}%
                                     {-3.25ex\@plus -1ex \@minus
                                     -.2ex}%
                                     {1.5ex \@plus .2ex}%
                                     {\normalfont\bfseries}}

\begin{document}

\newcommand{\bea}{\begin{eqnarray}}
\newcommand{\eea}{\end{eqnarray}}
\newcommand{\eeq}{\end{equation}}
\newcommand{\beq}{\begin{equation}}
\newcommand{\eg}{\emph{e.g.}\;}
\newcommand{\ie}{\emph{i.e.}\;}
\newcommand{\etal}{\emph{et al.}}
\newcommand{\nn}{\nonumber}
\newcommand{\pl}{\partial}
\newcommand{\SO}{\mathrm{SO}}
\newcommand{\SU}{\mathrm{SU}}
\newcommand{\U}{\mathrm{U}}

\makeatletter
\@addtoreset{equation}{section}
\makeatother
\renewcommand{\theequation}{\thesection.\arabic{equation}}

\rightline{QMUL-PH-08-14} \rightline{Imperial-TP-08-JB-01}
\vspace{2truecm}

\vspace{15pt}

%%%%%%%%%%%%%%%%%

\centerline{\LARGE \bf A note on Quantum Aspects of Multiple Membranes} \vspace{1truecm}
\thispagestyle{empty} \centerline{
    {\large \bf James Bedford${}^{a,b,}$}\footnote{E-mail:
                                  {\tt j.bedford@imperial.ac.uk}}
    {\bf and}
    {\large \bf David Berman${}^{c,}$}\footnote{E-mail:
                                  {\tt d.s.berman@qmul.ac.uk}}
                                                       }

\vspace{.4cm}
\centerline{{\it ${}^a$ Theoretical Physics Group}}
\centerline{{\it The Blackett Laboratory, Imperial College London}}
\centerline{{\it Prince Consort Road, London SW7 2AZ, UK}}

\vspace{.4cm}
\centerline{{\it ${}^b$ Institute for Mathematical Sciences}} \centerline{{\it Imperial College London}}
\centerline{{\it 53 Prince's Gate, London SW7 2PG, UK}}

\vspace{.4cm}
\centerline{{\it ${}^c$ Centre for Research in String Theory, Department of Physics}}
\centerline{{ \it Queen Mary, University of London}} \centerline{{\it Mile End Road, London E1 4NS, UK}}

\vspace{1.5truecm}

%%%%%%%%%%%%%%%%%
\thispagestyle{empty}

\centerline{\bf ABSTRACT}

\vspace{.5truecm}

\noindent In this note we investigate quantum aspects of the newly proposed theory of
multiple membranes put forward by Bagger and Lambert. In particular we analyse the possibility of a
finite renormalisation of the coupling at one loop.

\vspace{.5cm}

\setcounter{page}{0}

\setcounter{footnote}{0}

\newpage

%---------------------------------------------------------------------------

\section{Introduction}

The theory of multiple membranes has been something of a mystery for
many years \cite{berman1}. Now, it appears real progress has been made
with a newly proposed theory by Bagger, Lambert and Gustavsson
\cite{BL1,BL2,BL3,Gus1}. This breakthrough immediately attracted significant
attention with a great deal of work exploring numerous
avenues \cite{everyone}. One of the key elements of the Bagger--Lambert
theory is the presence of a coupling given by the inverse of the
level, $k$ of the Chern--Simons theory. At large $k$ the theory becomes
perturbative. Clearly, a tunable dimensionless parameter is not to be expected from a
theory of membranes since by definition M--theory does not contain any
free parameters. This parameter therefore encodes a property of a
particular background. Its interpretation is that the background is
formed with a particular modding out by a $\mathbb{Z}_k$ action and
is discussed in \cite{tong,mukhi,malda} (see also \cite{Hanany:2008qc}). We will
happily accept the presence of a perturbative parameter in the
theory and without looking the gift horse in the mouth proceed
to use it to examine some quantum aspects of the theory.

An obvious question is whether the beta function
vanishes to give quantum consistency. In fact, through
work by Kapustin and Pronin \cite{kapustin}\footnote{We would like to thank Costis Papageorgakis
for bringing \cite{kapustin} to our attention.} on properties of Chern--Simons theories
coupled to matter this question may be immediately answered and
indeed the beta function must vanish.
(This has also been addressed directly at one loop by Gustavsson
\cite{Gus2}). Here, we will be concerned with the possibility of
a finite shift in the level, $k$ at one loop. Pure Chern--Simons
theories are known to produce such a shift in the
coupling at one loop once a careful regularisation is used
(see \eg \cite{AG,semenoff}), while supersymmetric Chern--Simons
theories have also been
investigated \cite{kao} with the possible one loop shift explored
in detail for a variety of different supersymmetries. (The fermion
content is crucial since integrating out massive
fermions is known to contribute to the shift through their effective
action \cite{redlich}).

One must interpret this carefully especially since the effect seems
scheme dependent. We will follow the view espoused in \cite{jonespoly}
for pure Chern--Simons where the shift could be seen from a careful
treatment of the
phase of the partition function. In \cite{jonespoly}, the partition function
of pure Chern--Simons was calculated nonperturbatively, where possible, and it was found to be a function of
the shifted level. This indicated that although the shift may be
derived at one loop, the calculation is picking up that the full
nonperturbative result will be a function not of $k$ but of the shifted $k$.
In the present scenario, the most immediate physical effect caused by such a shift will be on
the moduli space which depends critically on $k$ \cite{tong,mukhi,malda}.

\section{Bagger--Lambert Theory}

We now describe the theory of Bagger, Lambert and Gustavsson \cite{BL1,BL2,BL3,Gus1} using the
conventions of Van Raamsdonk \cite{raamsdonk}. There are
eight scalars $X^I$, $I=1\ldots 8$ valued in $\SO(4)$ or equivalently the
bifundamental of $\SU(2)\times\SU(2)$ as follows: $X^I={1 \over 2} X^I_a
\sigma^a$ where $\sigma^a=(i \sigma^i, 1)$ and $\sigma^i$ are the
Pauli matrices. (In what follows $a,b,c=1\ldots 4$ and $i,j,k=1\ldots 3$). There are
eight fermions and their conjugates similarly valued in $\SO(4)$ and two gauge fields $A_\mu$
and ${\hat{A}}_\mu$ valued in $\SU(2)$ \ie\ $A_\mu = A^+_{\mu\,i} \sigma^i$
and ${\hat{A}}_\mu = A^-_{\mu\,i} \sigma^i$, where $A^+$ and $A^-$ are the self--dual and
anti--self--dual parts of the $\SO(4)$ gauge field respectively.
The gauge fields couple to matter through the covariant
derivative:
\beq
D_\mu X^I=\pl_\mu X^I +iA_\mu X^I -i X^I {\hat{A}}_\mu \, ,
\eeq
and the action is given by:
\bea
\label{action}
S &=& \int\! {\rm d}^3 x\ {\rm{tr}}\Big[ (D^\mu X^I)^\dagger D_\mu X^I +i
  {\bar{\psi}}^\dagger \Gamma^\mu D_\mu \psi \nn\\
  &-& {1 \over 3}\cdot {4\pi\over k}\,i\, {\bar{\psi}}^\dagger \Gamma_{IJ} \left( X^I (X^J)^\dagger +X^J
  \psi^\dagger X^I + \psi (X^I)^\dagger X^J \right) \nn\\
  &-& {2 \over 3}\cdot \left(4\pi\over k\right)^2
  X^{[I}X^{J\dagger}X^{K]}X^{K\dagger}X^JX^{I\dagger} \nn\\
  &+& {k \over 4\pi} \epsilon^{\mu \nu \lambda} (A_\mu \pl_\nu A_\lambda +
  {2\over3} i A_\mu A_\nu A_\lambda) - {k \over 4\pi} \epsilon^{\mu \nu
    \lambda}( {\hat{A}}_\mu \pl_\nu {\hat{A}}_\lambda +
  {2\over3} i{\hat{A}}_\mu {\hat{A}}_\nu {\hat{A}}_\lambda )\Big] \ .
\eea
The first line contains the usual kinetic terms, the second the Yukawa couplings,
the third the sextic interaction and final line the two Chern--Simons
actions for the vector potentials.

As described in \cite{gomis} we may also consider a massive deformation
of the theory that still preserves all the supersymmetries. This is
given by adding the following term to the action\footnote{In fact, one must also
add a mass--dependent potential in order for supersymmetry to be preserved, but this
will not play any role in our discussions.}:
\beq
\label{massdef}
S_{\textrm{mass}}= \int\!\!{\rm d}^3x\,\left(- \mu^2\, {\rm tr}\,( X^I X^I) + i\mu\, {\rm tr}\,(\bar{\psi}
\Gamma_{3456} \psi)\right) \, .
\eeq

Na\"{\i}vely one would expect the mass term for fermions to affect the one loop shift in
$k$, \cite{redlich,kao}. However, as we shall see the presence of $\Gamma_{3456}$ will
mean that the mass deformation will actually leave the one loop shift of $k$
invariant. That is, the shift will be independent of the fermion mass
deformation that preserves supersymmetry.

\section{One loop shift in the level}

We begin by reviewing the known arguments for a perturbative shift in
the coupling in light of Bagger--Lambert theory and go on to elucidate the effects of the extra
structure present. Primarily we follow the arguments in \cite{semenoff,kao} where careful treatment of the one
loop correction can be found. There is also an
excellent description of how this shift arises when being careful with the
phase of the path integral and the associated introduction of the eta
invariant via the Atiyah Patodi Singer index theorem
\cite{jonespoly}. When the partition function can be calculated
exactly it is then a function of the shifted $k$. This is similar to various 1+1 $d$ integrable models
where a finite shift in parameters makes the WKB approximation exact \cite{jonespoly}.
This second approach
provides insight into the nature of the shift and faith that the one
loop correction is something physical, but is not directly
extendible to the case of Chern--Simons coupled to matter that is required here.

To start with, we must identify an appropriate parameter with which to do perturbation theory.
Fortunately, as mentioned previously, the Bagger--Lambert action does contain such a parameter;
it is the Chern--Simons coefficient, $k$, which we will take to be large in order that there may
be a perturbative regime. One might make the objection that since $k$ must be integer-valued in order that
gauge invariance is preserved, we shouldn't really use it as a coupling constant since it cannot be
continuously varied. However, we will take a leaf out of earlier work on perturbative Chern--Simons theory
and brush this subtlety aside. Indeed, as we shall see, an important result is that quantum corrections
impose that $k$ can change by addition of an integer, thus preserving large gauge invariance.

Consider, then, one of the Chern--Simons terms in (\ref{action}). To regulate divergences, we
introduce a Yang--Mills term $-F^2/(2g_{\rm YM}^2)$. In three dimensions, $g_{\rm YM}^2$ has dimensions of mass which
makes the gauge fields topologically massive. In order to obtain physical results, however, we'll want to take
$g_{\rm YM}^2\rightarrow\infty$ which decouples the Yang--Mills regulator. For this purpose, it is useful to define the
dimensional parameter $m = g_{YM}^2(k/4\pi)$ and consider the limit as $m\rightarrow\infty$.
In order to perform calculations we must also fix a
gauge, which we take to be $\partial_{\mu}A^\mu=0$. This we do in the usual way by adding ghosts ($c,\,\bar{c}$) and the gauge-fixing
term $-(\partial_{\mu}A^{\mu})^2/\alpha$ to the action. Furthermore, we choose to work in Landau gauge where $\alpha=0$,
which has the dual advantages of radically simplifying the gluon propagator and taming infrared divergences
(see \eg \cite{semenoff} and references therein).

Introducing renormalisation functions in the usual way ($Z_{\bar{c}Ac}$ for the ghost--gluon vertex, $Z_c$ for the ghost
kinetic term and $Z_{\rm a}$ for the antisymmetric part of the gluon kinetic term) and a Ward identity associated with the ghost--gluon vertex,
we can then write the renormalised Chern--Simons coupling as
\beq
\label{couplingren}
k'=k\frac{Z_{\bar{c}Ac}^2}{Z_{\rm a}Z_c^2}\ .
\eeq
So, in order to determine the effect of renormalisation on the Chern--Simons
coefficient we need to examine the ghost self--energy, ghost--gluon interaction and the gluon
self--energy.

\subsection{Gluon self-energy}

In a general gauge, the classical $A^+$ propagator is given by
\bea
\label{fullprop}
\Delta_{\mu\nu}=-\frac{4\pi i}{k}\frac{m}{p^2\left(p^2-m^2\right)}
\left\{im\epsilon_{\mu\nu\lambda}p^{\lambda}+\eta_{\mu\nu}p^2
-p_{\mu}p_{\nu}\left(1-\frac{\alpha k}{4\pi m}\,\frac{p^2-m^2}{p^2}\right)\right\}\ ,
\eea
which, in the $m\rightarrow\infty$ limit (and in Landau gauge) reduces simply to
\beq
\Delta_{\mu\nu}=-\frac{4\pi}{k}\frac{\epsilon_{\mu\nu\lambda} p^{\lambda}}{p^2}
\eeq
as expected. The propagator for $A^-$ is simply minus this.

For the renormalisation functions here, the symmetries of the theory can be used to separate the gluon self-energy,
$\Pi_{\mu\nu}^{(1)}$ (at one loop),
into symmetric $(\Pi_{\rm s}^{(1)})$ and anti--symmetric $(\Pi_{\rm a}^{(1)})$ parts respectively:
\beq
\label{self-energy}
\Pi_{\mu\nu}^{(1)}=\Pi_{\rm a}^{(1)}\epsilon_{\mu\nu\lambda}p^{\lambda}
+\frac{1}{m}\Pi_{\rm s}^{(1)}(\eta_{\mu\nu}p^2-p_\mu p_\nu)\ ,
\eeq
where $Z_{\rm a}=1-\Pi_{\rm a}^{(1)}$ at one loop.

\subsubsection{Gluonic contributions}

In the case of pure Chern--Simons theory one can only have gluons and ghosts running in the loop via diagrams
(a)--(c) in Figure 1 and this was covered in detail in \cite{semenoff}. Furthermore, the absence of an
$A^+\!\!-\!A^-$ propagator means that these arguments apply separately to the self-dual and anti-self-dual parts
of the gauge field respectively. In this respect we can view the levels of the two Chern--Simons theories as
being essentially independent and see what effect the renormalisation has on each in turn.

\begin{figure}[ht]
\begin{center}
\begin{picture}(450,160)(0,0)
\SetOffset(5,0)
\put(0,10){
\SetWidth{0.8}
\Photon(20,20)(50,20){1.5}{5}
\Photon(120,20)(90,20){1.5}{5}
\Text(70,-20)[]{(d)}
\SetColor{Blue}
\DashCArc(70,20)(20,0,180){6}
\DashCArc(70,20)(20,180,360){6}
\SetColor{Black}
\Vertex(50,20){1.5}
\Vertex(90,20){1.5}
}
\put(140,-10){
\SetWidth{0.8}
\Photon(20,20)(120,20){1.5}{16.5}
\Text(70,0)[]{(e)}
\SetColor{Blue}
\DashCArc(70,42)(20,-90,270){6}
\SetColor{Black}
\Vertex(70,21.5){1.5}
}
\put(280,10){
\SetWidth{0.8}
\CCirc(70,20){20}{BrickRed}{White}
\Photon(20,20)(50,20){1.5}{5}
\Vertex(50,20){1.5}
\Vertex(90,20){1.5}
\Photon(120,20)(90,20){1.5}{5}
\Text(70,-20)[]{(f)}
}
\put(0,100){
\SetWidth{0.8}
\Photon(20,20)(50,20){1.5}{5}
\Vertex(50,20){1.5}
\PhotonArc(70,20)(20,0,180){1.5}{10}
\PhotonArc(70,20)(20,180,360){1.5}{10}
\Vertex(90,20){1.5}
\Photon(120,20)(90,20){1.5}{5}
\Text(70,-20)[]{(a)}
}
\put(140,80){
\SetWidth{0.8}
\Photon(20,20)(120,20){1.5}{16.5}
\Vertex(70,21.5){1.5}
\PhotonArc(70,44)(20,0,360){1.5}{19}
\Text(70,0)[]{(b)}
}
\put(280,100){
\SetWidth{0.8}
\Photon(20,20)(50,20){1.5}{5}
\Photon(120,20)(90,20){1.5}{5}
\Text(70,-20)[]{(c)}
\SetColor{OliveGreen}
\DashCArc(70,20)(20,0,180){2}
\DashCArc(70,20)(20,180,360){2}
\SetColor{Black}
\Vertex(90,20){1.5}
\Vertex(50,20){1.5}
}
\end{picture}
\end{center}
\label{gluon-self-energy-diagrams}
\caption{\emph{Processes contributing to the gluon self--energy at one loop with $F^2$ regularisation. (a) \& (b) are
corrections from virtual gluons; (c) is due to ghosts; (d) \& (e) are the exchange of virtual scalars, while (f) is the
fermionic contribution.}}
\end{figure}
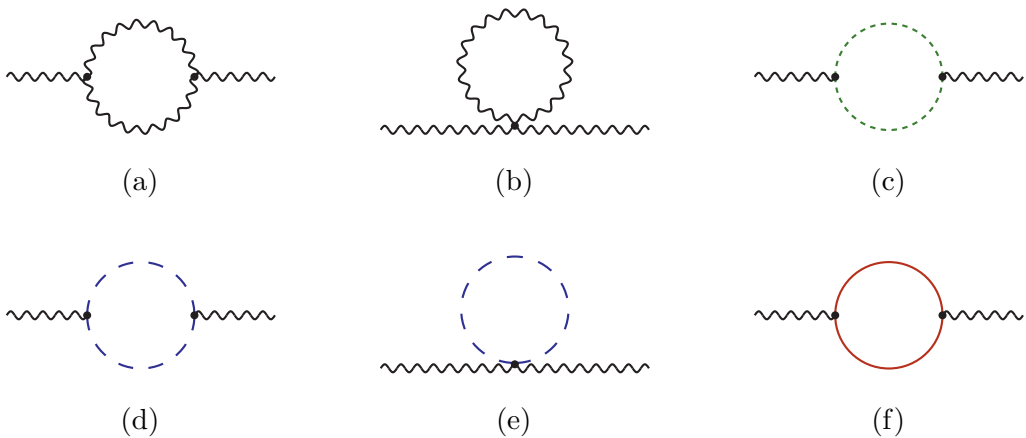

The calculations performed in \cite{semenoff} are therefore unchanged in this scenario and we
refer the reader to that paper for the full details. Here we simply present the results:
$\Pi_{\rm a}^{(1)}$ has a leading term proportional to $m/|m|={\rm sgn}(m)$.
This can be related to ${\rm sgn}(k)$, and when the contribution is evaluated in the $m\rightarrow\infty$ limit
one obtains\footnote{Note that in our conventions, $f^{acd}f^{bcd}=C_2\delta^{ab}$ so that $C_2(\SU(N))=N$.}
\beq
\Pi_{\rm a}^{(1)}=\frac{7}{3k}\,C_2\,{\rm sgn}(k)\ .
\eeq

\subsubsection{Matter contributions}

Of course, in Bagger--Lambert theory we also have contributions from matter fields as shown in
(d)--(f) of Figure 1 and it is important to ascertain how they contribute to the
gluon self--energy. In this context it is important to note that, as is well known \cite{semenoff},
$F^2$ regularisation does not regulate \emph{all} the divergences in one-loop diagrams. As such,
one may use a supplementary regularisation scheme. Indeed, as far as the evaluation of integrals
is concerned, dimensional continuation was already used to deal with those arising from
the gluonic contributions. For the matter fields, we may be tempted to add a supersymmetrised $F^2$ term
as a regulator in a similar spirit to the treatment of supersymmetric Yang--Mills-Chern--Simons
theories in \eg \cite{kao,witten-index}. However, this doesn't seem to make much sense as we already
have standard kinetic terms for the fermions and scalars. Similarly, we could conceivably add mass--terms
for the matter fields independently of the $F^2$ term. This is more like a regulator in the
infrared and in any case wouldn't give any contributions to the quantities of interest as we remove it.
Thus, we will content ourselves with dealing with any remaining divergences using a dimensional
continuation of the integrals where necessary\footnote{In fact, none of the integrals
involved pose any problems and we can happily evaluate them in $d$ dimensions and simply set $d=3$.}.

To begin with, scalars contribute via diagrams (d) and (e) of Figure 1 and it is easily verified that
they do not contribute to the antisymmetric part of $\Pi_{\mu\nu}$. A more generic way to view this is
that scalars are not parity violating. Secondly, fermions run in the loop via the last diagram, (f) of Figure 1, and
we would like to look at their contribution to $\Pi_{\rm a}^{(1)}$. For these purposes it is enough to look at the
momentum structure involved.

Each fermion vertex comes with a factor of $\Gamma_{\mu}$, while the fermion propagators take the usual
form $\Gamma_{\mu}p^\mu/p^2$. This means that after a little algebra
\beq
{\rm (f)}\propto \int\!\! {\rm d}^3q
\,\frac{q_{\mu}(q-p)_{\nu}+q_{\nu}(q-p)_{\mu}-\eta_{\mu\nu}q\cdot (q-p)}{q^2(q-p)^2}\ ,
\eeq
which is manifestly symmetric under the interchange of $\mu$ and $\nu$. The massless fermions in Bagger--Lambert
thus do not contribute to the antisymmetric part of $\Pi_{\mu\nu}$ at one loop.

Massive fermions, on the other hand, \emph{are} known to contribute to a shift in $k$, so it is interesting
to note that even if we were to add a standard mass term such as $m_{\psi}\bar{\psi}\psi$ to help regulate the fermions,
then these contributions vanish as we remove the regulator $m_{\psi}\rightarrow 0$. This is due to the fact that
the antisymmetric part of (f) generated by this mass is schematically $m_{\psi}+\mathcal{O}(m_{\psi}^2)$, which
vanishes as $m_{\psi}\rightarrow 0$. This is in accordance with expectations \cite{redlich}.

In conclusion, we can see that the matter fields do not contribute to the antisymmetric part of
the gluon self--energy at one loop.

\subsection{Ghost corrections}

Now we must look at the contributions to the ghost self--energy and the ghost--gluon interaction.
Since the ghosts do not couple to the matter fields it is obvious that the matter fields do not
contribute to the ghost self--energy or the ghost--gluon interaction at all at one loop. Thus
the one loop contributions to these quantities are precisely the same as in pure Chern--Simons theory
(with $F^2$ regularisation), a case which has previously been investigated in detail in \cite{semenoff}.
Thus we again spare the reader the details (referring instead to \cite{semenoff}) and present the results here.

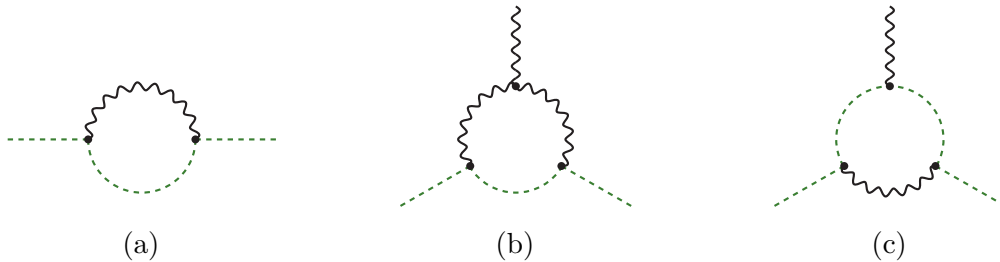
\begin{figure}[ht]
\begin{center}
\begin{picture}(450,80)(0,0)
\SetOffset(5,0)
\put(0,10){
\SetWidth{0.8}
\Text(70,-20)[]{(a)}
\PhotonArc(70,20)(20,0,180){1.5}{10}
\SetColor{OliveGreen}
\DashLine(20,20)(50,20){2}
\DashLine(120,20)(90,20){2}
\DashCArc(70,20)(20,180,360){2}
\SetColor{Black}
\Vertex(50,20){1.5}
\Vertex(90,20){1.5}
}
\put(140,10){
\SetWidth{0.8}
\Photon(70,40)(70,70){1.5}{5}
\Text(70,-20)[]{(b)}
\SetColor{OliveGreen}
\DashCArc(70,20)(20,210,330){2}
\DashLine(87,10)(113,-5){2}
\DashLine(53,10)(27,-5){2}
\SetColor{Black}
\PhotonArc(70,20)(20,-30,210){1.5}{14}
\Vertex(70,40){1.5}
\Vertex(87,10){1.5}
\Vertex(53,10){1.5}
}
\put(280,10){
\SetWidth{0.8}
\Photon(70,40)(70,70){1.5}{5}
\Text(70,-20)[]{(c)}
\SetColor{OliveGreen}
\DashCArc(70,20)(20,-30,210){2}
\DashLine(87,10)(113,-5){2}
\DashLine(53,10)(27,-5){2}
\SetColor{Black}
\PhotonArc(70,20)(20,210,330){1.5}{7}
\Vertex(70,40){1.5}
\Vertex(87,10){1.5}
\Vertex(53,10){1.5}
}
\end{picture}
\end{center}
\label{ghostdiagram}
\caption{\emph{Processes contributing to the ghost self--energy and the ghost--gluon vertex at one loop.}}
\end{figure}

The relevant diagrams are given in Figure 2, and there is only one contribution to the ghost
self--energy - that of Figure 2 (a). This evaluates to give
\beq
\Pi_c^{(1)}=-\frac{2}{3k}C_2\,{\rm sgn}(k)\ .
\eeq
In terms of the $\bar{c}Ac$ vertex corrections, many parts of the diagrams in Figure 2 (b) and (c)
cancel against each other and the remainder vanishes as $m\rightarrow\infty$ giving $Z_{\bar{c}Ac}=1$.
This is in any case expected from general arguments \cite{taylor}.

As we can see, the matter fields in the game do not contribute to the finite renormalisation of
$k$ and with $Z_c=1-\Pi_c^{(1)}$ at one loop we get
\bea
k'&=&k(1+\Pi_{\rm a}^{(1)}+2\Pi_c^{(1)})\nn\\
&=& k+C_2\,{\rm sgn}(k)\nn\\
&=& k+2\,{\rm sgn}(k)\ ,
\eea
where we have used that $C_2(\SU(2))=2$.
Note the crucial presence of ${\rm{sgn}}(k)$. In Bagger--Lambert theory we have two $\SU(2)$
theories with opposite levels, $k$ and $-k$. The fact that the
correction depends on the sign of $k$ means that the one loop
corrections preserve this structure, \ie the new levels are $k'$ and
$- k'$. Without this the SO(4) structure would be anomalous.

\section{Mass Deformation}

We may also consider the mass--deformed version of Bagger--Lambert
theory described in Section 2 which (when taken together with a mass--dependent potential) preserves $\mathcal{N}\!=\!8$
supersymmetry \cite{gomis}. As far as the quadratic terms go, this involves the addition of
a term $S_{\rm mass}$ given in (\ref{massdef}) to the action. The mass--dependent
potential is of fourth order in the scalars and therefore does not contribute to the
quantities of interest to us at one loop.

It is again clear that the scalar mass term cannot change the contributions to $\Pi_{\rm a}^{(1)}$ since
the scalars can still only give symmetric contributions and
don't affect the ghosts at this level. However, now that we have a mass term for the
fermions, they \emph{can} in principle contribute. The reason is that the fermion
propagator is now schematically of the form
\beq
\frac{\Gamma_\mu p^\mu+\mu\Gamma_{3456}}{p^2-\mu^2}\ ,
\eeq
and the terms arising from Figure 1 (f) which are proportional to $\mu$ contain
an odd number of $\Gamma_\mu$\footnote{Recall that the gamma matrices are split into
$\Gamma_{\mu}$ for $\mu=0,1,2$ and $\Gamma_{A}$ for $A=3\ldots 10$.} and can therefore contribute to $\Pi_{\rm a}^{(1)}$.
Nonetheless, as we see below, the contribution does in fact vanish.

A simple way to see this is that $\Gamma_{3456}$ squares to one and is traceless. Its
eigenvalues are thus equal numbers of $\pm 1$ and so the mass deformation of
(\ref{massdef}) is like adding equal numbers of fermions with mass $\mu$ as with mass $-\mu$. Since
their antisymmetric contribution to the gluon self--energy is proportional to their mass,
these contributions cancel out. It is reassuring to note that this is in accordance with
general arguments which give contributions to $k'$ proportional to ${\rm sgn}(\mu)$ \cite{redlich}.
The presence of equal numbers of oppositely--signed massive fermions is thus expected to give
no overall contribution.

From a more covariant point of view, where the fermions are still packaged into the single
spinor $\psi$, the antisymmetric part of Figure 1 (f) is easily calculated to be
\beq
{\rm (f)_a}\propto {\rm tr}\left(\Gamma_{\mu}\Gamma_{\nu}\Gamma_{\lambda}\Gamma_{3456}\right)\ .
\eeq
By Lorentz invariance this can only be proportional to $\epsilon_{\mu\nu\lambda}$
and in order to fix the constant of proportionality we can consider the case of
$\mu=0,\nu=1,\lambda=2$. By considering the relation $\Gamma_{0123456789(10)}=-1\!\!1$
we can see that
\beq
\Gamma_{0123456}=\pm\Gamma_{789(10)}\ ,
\eeq
depending on the signature of spacetime, and since ${\rm tr}\,\Gamma_{789(10)}=0$ it is
clear that the constant of proportionality is just zero. Thus these massive fermions do
not contribute to a shift in $k$.\footnote{It is interesting to note here that we could
have used this mass deformation as a supersymmetric regulator for the matter fields
and taken $\mu\rightarrow 0$ at the end of the day. Of course this would give the same results as
previously found in Section 3.} It is quite satisfying that the
supersymmetry preserving mass deformation leaves $k$ invariant even though a
canonical mass deformation would certainly lead to a different shift
in the one loop correction to $k$.

In conclusion, in both the original Bagger--Lambert theory and the deformed version, one
expects one loop quantum corrections to shift the coupling by two:
\beq
k\rightarrow k+2\,{\rm sgn}(k)\ .
\eeq

\section{Discussion}

In the proposed more general theory of \cite{malda} (see also \cite{Benna:2008zy}) with only
${\cal{N}}=6$ supersymmetry manifest there is $\U(N)\times \U(N)$ bifundamental
matter coupled to Chern--Simons. In that theory,
the link to Bagger--Lambert is that for $N\!=\!2$ the theory is expected
to have extra symmetries which promote the $\mathcal{N}\!=\!6$ supersymmetry to
$\mathcal{N}\!=\!8$. However, as a starting point for the $\mathcal{N}\!=\!6$
theory one may take $\mathcal{N}\!=\!4$ super--Yang--Mills plus Chern--Simons terms and
integrate out the massive fields. In doing so, as discussed
in \cite{malda} the fermions would cause a shift in $k$. This shift would then in
that theory be subsequently cancelled by the shift in $k$ due to the one
loop correction from the Chern--Simons field as described above. Thus, overall there
would be no shift in $k$. From this perspective the Bagger--Lambert
theory as written above would be an effective theory where one loop
effects have already been included. One cannot say a priori whether this
is correct though there is now more evidence as to the success of
\cite{malda} with \cite{minwalla}.

This paper took the approach of looking at the possible shift in $k$ from
a perturbative point of view using regularisation by addition of regulator terms
such as the Yang--Mills term. In this context it does not seem to be possible to
regularise the theory in the UV in a supersymmetric way. However, there may be
a sense in which it \emph{is} possible to regulate the theory while preserving at least
some of the supersymmetry: One could consider regulating by replacing the entire Bagger--Lambert
action with a supersymmetric YM--CS action of the sort encountered in \cite{kao}\footnote{We would like to thank Seok Kim for illuminating
discussions on this point.}. As long
as the specific form of this action preserves $\mathcal{N}\geq 2$ supersymmetries\footnote{Though it doesn't seem
likely that we could preserve the full $\mathcal{N}\!=\!8$ SUSY, at least explicitly.} then
performing a one--loop renormalisation would lead to a cancellation between bosons and
fermions such that there is no overall shift in $k$ \cite{kao}. Removing the UV
regulator should be equivalent to integrating out the massive fields and thus one
would expect to recover Bagger--Lambert theory in this limit \`{a} la \cite{malda} and
without any shift in $k$.\footnote{Note that if it
is possible to do this procedure with a YM--CS action
preserving only $\mathcal{N}\!=\!1$ supersymmetry, which is not entirely clear, then one would still see a shift in $k$.}
On the other hand, it seems somewhat drastic to regulate by replacing the entire action with something new.

Thus, although the two approaches of \cite{BL1,BL2,BL3,Gus1} and of \cite{malda} are classically
equivalent, they may not be so at the quantum level. A standard regularisation by addition of
regulator terms as we have examined would seem to break the quantum equivalence and lead
to a one--loop shift in $k$ for BL. If one additionally makes the assumption that the
calculated shift is true for all $k$ (which we imagine to be the case but certainly cannot
derive given the perturbative nature of these calculations), then the moduli space
at $k=1$ would become corrected which seems unphysical.

\pagebreak

\section*{Acknowledgements}

We would like to thank Nick Dorey, Seok Kim, Costis Papageorgakis,
Riccardo Ricci, Dan Thompson, Dave Tong and Gabriele Travaglini for
useful discussions. This work was in part supported by the EC Marie Curie
Research Training Network, MRTN-CT-2004-512194. J.B. is supported by
an STFC Research Fellowship.

\mbox{}


\begin{thebibliography}{99}

\bibitem{berman1}
  D.~S.~Berman,
  %``M-theory branes and their interactions,''
  Phys.\ Rept.\  {\bf 456} (2008) 89
  [arXiv:0710.1707 [hep-th]].
  %%CITATION = PRPLC,456,89;%%

\bibitem{BL1}
  J.~Bagger and N.~Lambert,
  ``Modeling multiple M2's,''
  Phys.\ Rev.\  D {\bf 75}, 045020 (2007)
  [arXiv:hep-th/0611108].
  %%CITATION = PHRVA,D75,045020;%%

\bibitem{BL2}
  J.~Bagger and N.~Lambert,
  ``Gauge Symmetry and Supersymmetry of Multiple M2-Branes,''
  Phys.\ Rev.\  D {\bf 77}, 065008 (2008)
  [arXiv:0711.0955 [hep-th]].
  %%CITATION = PHRVA,D77,065008;%%

\bibitem{BL3}
  J.~Bagger and N.~Lambert,
  ``Comments On Multiple M2-branes,''
  JHEP {\bf 0802}, 105 (2008)
  [arXiv:0712.3738 [hep-th]].
  %%CITATION = JHEPA,0802,105;%%

\bibitem{Gus1}
  A.~Gustavsson,
  %``Algebraic structures on parallel M2-branes,''
  arXiv:0709.1260 [hep-th].
  %%CITATION = ARXIV:0709.1260;%%

\bibitem{everyone}
S.~Mukhi and C.~Papageorgakis,
  ``M2 to D2,''
  JHEP {\bf 0805}, 085 (2008)
  [arXiv:0803.3218 [hep-th]];\newline
  M.~A.~Bandres, A.~E.~Lipstein and J.~H.~Schwarz,
  ``N = 8 Superconformal Chern--Simons Theories,''
  JHEP {\bf 0805}, 025 (2008)
  [arXiv:0803.3242 [hep-th]];\newline
    A.~Gustavsson,
  ``Selfdual strings and loop space Nahm equations,''
  JHEP {\bf 0804} (2008) 083
  [arXiv:0802.3456 [hep-th]];\newline
  D.~S.~Berman, L.~C.~Tadrowski and D.~C.~Thompson,
  ``Aspects of Multiple Membranes,''
  arXiv:0803.3611 [hep-th];\newline
  A.~Morozov,
  ``On the Problem of Multiple M2 Branes,''
  JHEP {\bf 0805}, 076 (2008)
  [arXiv:0804.0913 [hep-th]];\newline
  U.~Gran, B.~E.~W.~Nilsson and C.~Petersson,
  ``On relating multiple M2 and D2-branes,''
  arXiv:0804.1784 [hep-th];\newline
  P.~M.~Ho, R.~C.~Hou and Y.~Matsuo,
  ``Lie 3-Algebra and Multiple M2-branes,''
  JHEP {\bf 0806}, 020 (2008)
  [arXiv:0804.2110 [hep-th]];\newline
  E.~A.~Bergshoeff, M.~de Roo and O.~Hohm,
  ``Multiple M2-branes and the Embedding Tensor,''
  arXiv:0804.2201 [hep-th];\newline
  K.~Hosomichi, K.~M.~Lee and S.~Lee,
  ``Mass-Deformed Bagger-Lambert Theory and its BPS Objects,''
  arXiv:0804.2519 [hep-th];\newline
  G.~Papadopoulos,
  ``M2-branes, 3-Lie Algebras and Plucker relations,''
  JHEP {\bf 0805}, 054 (2008)
  [arXiv:0804.2662 [hep-th]];\newline
  J.~P.~Gauntlett and J.~B.~Gutowski,
  ``Constraining Maximally Supersymmetric Membrane Actions,''
  arXiv:0804.3078 [hep-th];\newline
  G.~Papadopoulos,
  ``On the structure of k-Lie algebras,''
  Class.\ Quant.\ Grav.\  {\bf 25}, 142002 (2008)
  [arXiv:0804.3567 [hep-th]];\newline
  P.~M.~Ho and Y.~Matsuo,
  ``M5 from M2,''
  arXiv:0804.3629 [hep-th];\newline
  J.~Gomis, G.~Milanesi and J.~G.~Russo,
  ``Bagger-Lambert Theory for General Lie Algebras,''
  JHEP {\bf 0806}, 075 (2008)
  [arXiv:0805.1012 [hep-th]];\newline
  S.~Benvenuti, D.~Rodriguez-Gomez, E.~Tonni and H.~Verlinde,
  ``N=8 superconformal gauge theories and M2 branes,''
  arXiv:0805.1087 [hep-th];\newline
  P.~M.~Ho, Y.~Imamura and Y.~Matsuo,
  ``M2 to D2 revisited,''
  arXiv:0805.1202 [hep-th];\newline
  A.~Morozov,
  ``From Simplified BLG Action to the First-Quantized M-Theory,''
  arXiv:0805.1703 [hep-th];\newline
  H.~Fuji, S.~Terashima and M.~Yamazaki,
  ``A New N=4 Membrane Action via Orbifold,''
  arXiv:0805.1997 [hep-th];\newline
  P.~M.~Ho, Y.~Imamura, Y.~Matsuo and S.~Shiba,
  ``M5-brane in three-form flux and multiple M2-branes,''
  arXiv:0805.2898 [hep-th];\newline
  C.~Krishnan and C.~Maccaferri,
  ``Membranes on Calibrations,''
  arXiv:0805.3125 [hep-th];\newline
  Y.~Song,
  ``Mass Deformation of the Multiple M2 Branes Theory,''
  arXiv:0805.3193 [hep-th];\newline
  I.~Jeon, J.~Kim, N.~Kim, S.~W.~Kim and J.~H.~Park,
  ``Classification of the BPS states in Bagger-Lambert Theory,''
  arXiv:0805.3236 [hep-th];\newline
  M.~Li and T.~Wang,
  ``M2-branes Coupled to Antisymmetric Fluxes,''
  arXiv:0805.3427 [hep-th];\newline
  K.~Hosomichi, K.~M.~Lee, S.~Lee, S.~Lee and J.~Park,
  ``N=4 Superconformal Chern-Simons Theories with Hyper and Twisted Hyper
  Multiplets,''
  arXiv:0805.3662 [hep-th];\newline
  S.~Banerjee and A.~Sen,
  ``Interpreting the M2-brane Action,''
  arXiv:0805.3930 [hep-th];\newline
  H.~Lin,
  ``Kac-Moody Extensions of 3-Algebras and M2-branes,''
  arXiv:0805.4003 [hep-th];\newline
  J.~Figueroa-O'Farrill, P.~de Medeiros and E.~Mendez-Escobar,
  ``Lorentzian Lie 3-algebras and their Bagger-Lambert moduli space,''
  arXiv:0805.4363 [hep-th];\newline
  F.~Passerini,
  ``M2-Brane Superalgebra from Bagger-Lambert Theory,''
  arXiv:0806.0363 [hep-th];\newline
  B.~Ezhuthachan, S.~Mukhi and C.~Papageorgakis,
  ``D2 to D2,''
  arXiv:0806.1639 [hep-th];\newline
  C.~Ahn,
  ``Holographic Supergravity Dual to Three Dimensional N=2 Gauge Theory,''
  arXiv:0806.1420 [hep-th];\newline
  S.~Cecotti and A.~Sen,
  ``Coulomb Branch of the Lorentzian Three Algebra Theory,''
  arXiv:0806.1990 [hep-th];\newline
  A.~Mauri and A.~C.~Petkou,
  ``An N=1 Superfield Action for M2 branes,''
  arXiv:0806.2270 [hep-th];\newline
  E.~A.~Bergshoeff, M.~de Roo, O.~Hohm and D.~Roest,
  ``Multiple Membranes from Gauged Supergravity,''
  arXiv:0806.2584 [hep-th];\newline
  P.~de Medeiros, J.~Figueroa-O'Farrill and E.~Mendez-Escobar,
  ``Metric Lie 3-algebras in Bagger-Lambert theory,''
  arXiv:0806.3242 [hep-th];\newline
  M.~Blau and M.~O'Loughlin,
  ``Multiple M2-Branes and Plane Waves,''
  arXiv:0806.3253 [hep-th];\newline
  T.~Nishioka and T.~Takayanagi,
  ``On Type IIA Penrose Limit and N=6 Chern-Simons Theories,''
  arXiv:0806.3391 [hep-th];\newline
  Y.~Honma, S.~Iso, Y.~Sumitomo and S.~Zhang,
  ``Scaling limit of N=6 superconformal Chern-Simons theories and Lorentzian
  Bagger-Lambert theories,''
  arXiv:0806.3498 [hep-th].

\bibitem{tong}
  N.~Lambert and D.~Tong,
  ``Membranes on an Orbifold,''
  arXiv:0804.1114 [hep-th].
  %%CITATION = ARXIV:0804.1114;%%

\bibitem{mukhi}
  J.~Distler, S.~Mukhi, C.~Papageorgakis and M.~Van Raamsdonk,
  ``M2-branes on M-folds,''
  JHEP {\bf 0805}, 038 (2008)
  [arXiv:0804.1256 [hep-th]].
  %%CITATION = JHEPA,0805,038;%%

\bibitem{malda}
  O.~Aharony, O.~Bergman, D.~L.~Jafferis and J.~Maldacena,
  ``N=6 superconformal Chern-Simons-matter theories, M2-branes and their
  gravity duals,''
  arXiv:0806.1218 [hep-th].
  %%CITATION = ARXIV:0806.1218;%%

\bibitem{Hanany:2008qc}
  A.~Hanany, N.~Mekareeya and A.~Zaffaroni,
  ``Partition Functions for Membrane Theories,''
  arXiv:0806.4212 [hep-th].
  %%CITATION = ARXIV:0806.4212;%%

\bibitem{kapustin}
  A.~N.~Kapustin and P.~I.~Pronin,
  ``Nonrenormalization theorem for gauge coupling in (2+1)-dimensions,''
  Mod.\ Phys.\ Lett.\  A {\bf 9}, 1925 (1994)
  [arXiv:hep-th/9401053].
  %%CITATION = MPLAE,A9,1925;%%

\bibitem{Gus2}
  A.~Gustavsson,
  ``One-loop corrections to Bagger-Lambert theory,''
  arXiv:0805.4443 [hep-th].
  %%CITATION = ARXIV:0805.4443;%%

\bibitem{AG}
  L.~Alvarez-Gaume, J.~M.~F.~Labastida and A.~V.~Ramallo,
  ``A Note on Perturbative Chern-Simons Theory,''
  Nucl.\ Phys.\  B {\bf 334}, 103 (1990).
  %%CITATION = NUPHA,B334,103;%%

\bibitem{semenoff}
  W.~Chen, G.~W.~Semenoff and Y.~S.~Wu,
  ``Two loop analysis of nonAbelian Chern-Simons theory,''
  Phys.\ Rev.\  D {\bf 46}, 5521 (1992)
  [arXiv:hep-th/9209005].
  %%CITATION = PHRVA,D46,5521;%%

\bibitem{kao}
  H.~C.~Kao, K.~M.~Lee and T.~Lee,
  ``The Chern-Simons coefficient in supersymmetric Yang-Mills Chern-Simons
  theories,''
  Phys.\ Lett.\  B {\bf 373}, 94 (1996)
  [arXiv:hep-th/9506170].
  %%CITATION = PHLTA,B373,94;%%

\bibitem{redlich}
  A.~N.~Redlich,
  ``Parity Violation And Gauge Noninvariance Of The Effective Gauge Field
  Action In Three-Dimensions,''
  Phys.\ Rev.\  D {\bf 29}, 2366 (1984).
  %%CITATION = PHRVA,D29,2366;%%

\bibitem{jonespoly}
  E.~Witten,
  ``Quantum field theory and the Jones polynomial,''
  Commun.\ Math.\ Phys.\  {\bf 121}, 351 (1989).
  %%CITATION = CMPHA,121,351;%%

\bibitem{raamsdonk}
  M.~Van Raamsdonk,
  ``Comments on the Bagger-Lambert theory and multiple M2-branes,''
  JHEP {\bf 0805}, 105 (2008)
  [arXiv:0803.3803 [hep-th]].
  %%CITATION = JHEPA,0805,105;%%

\bibitem{gomis}
  J.~Gomis, A.~J.~Salim and F.~Passerini,
  ``Matrix Theory of Type IIB Plane Wave from Membranes,''
  arXiv:0804.2186 [hep-th].
  %%CITATION = ARXIV:0804.2186;%%

\bibitem{witten-index}
  E.~Witten,
  ``Supersymmetric index of three-dimensional gauge theory,''
  arXiv:hep-th/9903005.
  %%CITATION = HEP-TH/9903005;%%

\bibitem{taylor}
  J.~C.~Taylor,
  ``Ward Identities And Charge Renormalization Of The Yang-Mills Field,''
  Nucl.\ Phys.\  B {\bf 33}, 436 (1971).
  %%CITATION = NUPHA,B33,436;%%

\bibitem{Benna:2008zy}
  M.~Benna, I.~Klebanov, T.~Klose and M.~Smedback,
  ``Superconformal Chern-Simons Theories and $AdS_4/CFT_3$ Correspondence,''
  arXiv:0806.1519 [hep-th].
  %%CITATION = ARXIV:0806.1519;%%

\bibitem{minwalla}
  J.~Bhattacharya and S.~Minwalla,
  ``Superconformal Indices for ${\cal N}=6$ Chern Simons Theories,''
  arXiv:0806.3251 [hep-th].
  %%CITATION = ARXIV:0806.3251;%%


\end{thebibliography}
\end{document}